\documentclass[12pt]{article}
\usepackage{verbatim}
\usepackage{latexsym}
\usepackage{amsfonts}
\usepackage{amsmath}
\usepackage{amsthm}
\usepackage{url}

\newtheorem{theorem}{Theorem}
\newtheorem{lemma}[theorem]{Lemma}
\newtheorem{corollary}[theorem]{Corollary}

\theoremstyle{definition}
\newtheorem{definition}{Definition}
\newtheorem{example}{Example}

\newcommand{\coeff}[1]{[z^{#1}]}

\begin{document}
\title{A tight upper bound on non-overlapping codes}
\author{Simon R. Blackburn\\
Mathematics, Royal Holloway University of London\\
Egham, Surrey TW20 0EX\\
\texttt{s.blackburn@rhul.ac.uk}\\
\\
\emph{Dedicated to Tuvi Etzion}\\
\emph{on the occasion of his 70th birthday.}}

\maketitle

\begin{abstract}
Non-overlapping codes (also known as cross bifix-free codes or strong comma-free codes) have now been studied for over 60 years as part of information theory, with regular new applications appearing and much recent interest. Let $X$ be a $q$-ary non-overlapping code of length $n$. The paper contains a self-contained proof that
\[
|X|\leq \left(\frac{q-1}{qe} +O(n^{-1/4})\right)\frac{q^n}{n}
\]
as $n\rightarrow\infty$ with $q$ fixed, where $e$ is the base of the natural logarithm. The constant $(q-1)/(qe)$ improves, by a factor of $(q-1)/q$, the best previously known bound due to Levenshtein (1970). Indeed, constructions of large non-overlapping codes show that this constant is best possible. The paper builds on work by Fimmel, Michel, Pirot, Sereni and Str\"ungmann (2019), and by Stanovnik, Mo\v skon and Mraz (2024).
\end{abstract}

\section{Introduction}
\label{sec:introduction}

Let $F$ be an alphabet of size $q$, where $q\geq 2$. Let $n$ be an integer, $n\geq 2$. Let $u,v\in F^n$ be words of length $n$ over $F$, and write $u=(u_1,u_2,\dots ,u_n)$ and $v=(v_1,v_2,\ldots,v_n)$. We say that $u$ and $v$ are \emph{overlapping} if there exists an integer $s$ with $1\leq s\leq n-1$ such that either $u_{n-s+i}=v_{i}$ for $1\leq i\leq s$, or  $u_{i}=v_{n-s+i}$ for $1\leq i\leq s$. For example, when $q=2$ and $n=5$ the words $u=001\underline{01}$ and $v=\underline{01}111$ are overlapping (with $s=2$), as their underlined portions overlap. However, the words $00101$ and $00111$ are non-overlapping. Note that $\underline{00}1\underline{00}$, for example, overlaps with itself.

A code $X\subseteq F^n$ is \emph{non-overlapping} if all pairs $u,v\in X$ of codewords (not necessarily distinct) are non-overlapping. For example, the code $X=\{00101,00111\}$ is not overlapping. Non-overlapping codes were introduced by Gilbert~\cite{Gilbert60} and Levenshtein~\cite{Levenshtein64} in the context of asynchronous data transmission: if data is transmitted by concatenating elements from a non-overlapping code, and part of the data stream is received without any information about the starting point of each codeword, then individual codewords can be unambiguously identified, and synchronisation can be established, because of the non-overlapping property: slide a window of length $n$ along the data stream until a codeword is reached. The importance of synchronisation, and applications to (for example) DNA sequencing, have meant that variants of non-overlapping codes have been actively studied for many years, and reinvented several times (hence the many names for the concept). See, for example,~\cite{Bajic,BajicStojanovic,BajicStojanovicLindner,BilottaGrazzini,BilottaPergola,Blackburn15,CheeKiah13,FimmelMichel19,FimmelMichel23,Levenshtein04,WijngaardenWillink,StanovnikMoskon24,Zakharov26}.

One early surprise in the subject is that non-overlapping codes can be large. The following binary construction, due to Gilbert~\cite{Gilbert60} and Levenshtein~\cite{Levenshtein70}, shows this. Fix a positive integer $k$ with $k\leq n-2$, and consider the code $C$ of all words of the form $0^k1w1$ where $w$ has length $n-k-2$ and is $0^k$-free (in the sense it has no $k$ consecutive terms all $0$). Then $C$ is non-overlapping, and when $k$ is chosen appropriately (in particular, to be logarithmic in $n$) we have
\[
|C|\sim \frac{1}{2e}\frac{2^n}{n}
\]
as $n\rightarrow\infty$ along some infinite sequence of integers. Thus the rate of the best binary non-overlapping codes approaches $1$. The natural generalisation of this construction to $q$-ary codes is also large. So, for an alphabet $F$ with $0\in F$ we take $C$ to be the set of all words of the form $0^k1w1$ where $w$ has length $n-k-2$ and is $0^k$-free, and we may calculate that
\[
|C|\sim \frac{q-1}{qe}\frac{q^n}{n}
\]
as $n\rightarrow\infty$ along some infinite sequence of integers.

How good are the constructions above? One positive answer was provided by Levenshtein~\cite{Levenshtein70}, who proved that a non-overlapping code of length $n$ has cardinality bounded above by
\begin{equation}
\label{eqn:Levenshtein}
\left(1-\frac{1}{n}\right)^{n-1}\frac{q^n}{n}=\left(\frac{1}{e}+O(n^{-1})\right)\frac{q^n}{n}
\end{equation}
as $n\rightarrow\infty$, where $e$ is the base to the natural logarithm. This bound shows that the construction above is within a factor of $q/(q-1)$ of optimality. This leaves the tantalising open question: For $q$ fixed, what is the best possible constant $c$ such that $|C|\leq (c+o(1))q^n/n$ for all $q$-ary non-overlapping codes $C$? The construction and bound above show that  $(q-1)/(qe)\leq c\leq 1/e$.

We remark that Levenshtein's upper bound is actually tight when $n$ divides $q$ (see Blackburn~\cite[Theorem~4]{Blackburn15}). Moreover, the upper bound remains good even when $n$ does not divide $q$, in the sense that there is a positive absolute constant $c$ such that there exists a $q$-ary non-overlapping code of length $n$ and cardinality at least $cq^n/n$ for all integers $q\geq 2$ and $n\geq 2$; see Blackburn~\cite[Theorem~8]{Blackburn15}.

Another answer to the question of whether the code $C$ above is good is provided by asking whether $C$ has the largest cardinality amongst all $q$-ary non-overlapping codes of length $n$. For example, restricting to binary codes for the moment, Chee \emph{et al}~\cite{CheeKiah13} computed the maximal size of a non-overlapping code for $n\leq 16$. This computation was extended by Stanovnik, Mo\v skon and Mraz to $n\leq 29$, by using work of Fimmel, Michel, Pirot, Sereni and Str\"ungmann~\cite{FimmelMichel19} to reframe the problem as an integer optimisation problem. Looking at these results, we get a second surprise: Even though $C$ (when $k$ is chosen to maximise $|C|$) has maximal cardinality for many values of $n$, there are codes with larger cardinality when $n=9$ and when $18\leq n\leq 22$. However, the gap between the cardinality of the largest non-overlapping code and the cardinality of the code $C$ above is not large, so it is very possible that $C$ has maximal cardinality for infinitely many values of $n$.

The purpose of this paper is to provide a self-contained proof of an improvement to Levenshtein's upper bound~\eqref{eqn:Levenshtein} from 1970, to provide a tight leading term when $n\rightarrow\infty$ with $q$ fixed. We will prove the following theorem.

\begin{theorem}
\label{thm:main}
Let $q$ be fixed, $q\geq 2$. Let $X$ be a $q$-ary non-overlapping code of length $n$. Then
\begin{equation}
\label{eqn:main}
|X|\leq \left(\frac{q-1}{qe} +O(n^{-{1/4}})\right)\frac{q^n}{n}
\end{equation}
as $n\rightarrow\infty$. 
\end{theorem}

We remark that the constant $\frac{q-1}{qe}$ cannot be replaced by any smaller constant, as the constructions $C$ above show. 

There are three key ingredients of the proof. First, there is the work of Fimmel \emph{et al}~\cite{FimmelMichel19}, who break up a maximal non-overlapping code into a collection of disjoint subsets of a simple form. Secondly, there is the insight by Stanovnik \emph{et al}~\cite{StanovnikMoskon24} that we may reduce the problem to providing an upper bound on the solutions of a certain integer maximisation problem. Finally, there is the bound itself, which relies (just like the original Levenshtein bound) on using some of the machinery from the theory of generating functions.

The rest of the paper is structured as follows.
In Section~\ref{sec:Fimmel} we restate and provide direct proofs of the properties of a certain decomposition of a maximal non-overlapping code due to Fimmel \emph{et al}. Some of the results in this section were proved in Stanovnik \emph{et al} (and were also strengthened in some cases). We end the section by showing that bounding a certain integer maximisation problem will give us an upper bound on non-overlapping codes; Stanovnik \emph{et al}~\cite[Proposition~10]{StanovnikMoskon24} provides a stronger result. In Section~\ref{sec:improved_upper}, we rephrase the integer maximisation problem in terms of a power series identity. This allows us to use techniques from the theory of generating functions to derive our upper bound, and finish the proof of Theorem~\ref{thm:main}.

\section{Decomposing codes, and an integer maximisation}
\label{sec:Fimmel}

We say that a non-overlapping code $X$ is \emph{maximal} if the only non-overlapping code $Y\subseteq F^n$ containing $X$ is the code $X$ itself. A non-overlapping code of maximal cardinality is maximal, but in fact the converse need not hold. 

In this section, we will provide a proof of the main theorem of Fimmel \emph{et~al}, and derive some immediate corollaries. In particular, following Stanovnik \emph{et al}, we reduce our upper bound to the study of an integer maximisation problem.

\begin{definition}
Let $X\subseteq F^i$ and $Y\subseteq F^j$ be codes of positive lengths $i$ and $j$ respectively. We say that the pair $(X,Y)$ is \emph{balanced} if the following property holds. For all $x\in X$ and $y\in Y$, if a non-empty suffix $\sigma$ of $x$ is equal to a (non-empty) prefix $\pi$ of $y$, then $\sigma=\pi\in\{x,y\}$.
\end{definition}
In other words, $(X,Y)$ is balanced if there are no partial overlaps between the end of $x$ and start of $y$, although a full overlap is allowed to occur. For example,
\begin{itemize}
\item the pair $(\{00100\},\{101\})$ is balanced (there are no overlaps);
\item the pair $(\{01111\},\{001\})$ is balanced (partial overlaps involving the end of $y\in Y$ and the start of $x\in X$ are fine);
\item $(\{11100\},\{110\})$ is balanced (so $y\in Y$ is allowed to occur in the middle of $x\in X$);
\item $(\{11111\},\{100\})$ is not balanced, as there are partial overlaps involving the end of $x\in X$ and the start of $y\in Y$.
\end{itemize}

We note that $(X,X)$ is balanced if and only if $X$ is non-overlapping. 

Let $X$ be a non-overlapping code of length $n$. Define the sets $L_i$, $R_i$ and $X_j$ for $1\leq i\leq n-1$ and $1\leq j\leq n$ as in~\cite{FimmelMichel19}. So we define $X_1=F$, and once $X_i$ is defined we set:
\begin{align*}
L_i&=\{w\in X_i: w\text{ is a prefix of a codeword in $X$}\},\\
R_i&=X_i\setminus L_i\text{ and}\\
X_{i+1}&=\bigcup_{a\in\{1,2,\ldots,i\}}L_aR_{i+1-a}.
\end{align*}
We define the \emph{closure} $\overline{X}$ of the non-overlapping code $X$ to be the code $X_n$.

\begin{example} Let $k$ be such that $1\leq k\leq n-2$ and let $F=\{0,1\}$. Define
\[
X=\{0^k1s1:s\text{ is a $0^k$-free sequence of length $n-k-2$}\}.
\]
We have:
\[
\begin{array}{ll}
L_1=\{0\} &R_1=\{1\}\\
L_{j}=\emptyset&R_j=\{0^{j-1}1\}\text{ for }2\leq j\leq k\\
L_{k+1}=\{0^k1\}&R_{k+1}=\emptyset\\
L_{k+j}=\{0^k1s1\}&R_{k+j}=\emptyset\text{ for }2\leq j\leq n-k-1,
\end{array}
\]
where in the last line the sequence $s$ varies over all sequences of length $j-2$ that are $0^k$-free. In this example, $\overline{X}=X$.
\end{example}

It is natural to think that if $X\subseteq Y$, where $X$ and $Y$ are non-overlapping codes, then $\overline{X}\subseteq \overline{Y}$. But this is not the case, as the following example shows. (This example also shows that $\overline{X}$ can strictly contain $X$.)

\begin{example}
Let $F=\{0,1\}$ and let $X=\{00101\}$. Then we may calculate
\[
\begin{array}{c|cc}
i&L_i&R_i\\\hline
1&\{0\}&\{1\}\\
2&\emptyset&\{01\}\\
3&\{001\}&\emptyset\\
4&\emptyset&\{0011\}
\end{array}
\]
and so $\overline{X}=\{00101,00011\}$. But consider the code $Y=\{00111,00101\}$. Then
\[
\begin{array}{c|cc}
i&L_i&R_i\\\hline
1&\{0\}&\{1\}\\
2&\emptyset&\{01\}\\
3&\{001\}&\emptyset\\
4&\{0011\}&\emptyset
\end{array}
\] 
and so $\overline{Y}=\{00111,00101\}=Y$. Note that $X\subseteq Y$ but $\overline{X}\not\subseteq \overline{Y}$ in this example.
\end{example}

\begin{theorem}
\label{thm:balanced}
Let $i$ and $j$ be integers such that $2\leq i,j\leq n$. The pair $(X_i,X_j)$ is balanced.
\end{theorem}

\begin{proof}
For $4\leq r\leq 2n$, let $P(r)$ be the statement that $(X_i,X_j)$ is balanced whenever $2\le i,j\le n$ and $i+j=r$. We will prove the theorem using induction on $r$.

When $r=4$, the statement $P(r)$ is trivially true as $X_i=X_2\subseteq L_1 R_1$ and $X_j=X_2\subseteq L_1R_1$ where $L_1$ and $R_1$ form a partition of $F$.

Fix a value $k>4$, and assume that $P(r)$ holds whenever $4\leq r<k$. We aim to show that $P(k)$ holds.

Let $i$ and $j$ be integers such that $i+j=k$ and $2\leq i,j\leq n$. We will show that $(X_i,X_j)$ is balanced. Let $w\in X_i$ and $w'\in X_j$, so there exist positive integers $i_0$, $i_1$, $j_0$, $j_1$ with $i_0+i_1=i$ and $j_0+j_1=j$ such that 
\begin{align*}
w&=w_0w_1\text{ for }w_0\in L_{i_0}, w_1\in R_{i_1}\text{ and }\\
w'&=w'_0w'_1\text{ for }w'_0\in L_{j_0}, w'_1\in R_{j_1}.
\end{align*}
Furthermore, suppose that there is a non-empty suffix $s$ of $w$ that is equal to a non-empty prefix $p$ of $w'$. To prove the inductive step, it suffices to show that $s\in\{w,w'\}$. For a word $v$ over $F$, we write $|v|$ for the length of $v$.

First assume that $|s|>|w_1|$. Then $w_0$ has a non-empty suffix $\sigma$ (of length $|s|-|w_1|$) that is a prefix of $w'$. When $i_0\geq 2$, since $w_0\in X_{i_0}$ and $w'\in X_j$, the statement $P(i_0+j)$ implies that $\sigma\in\{w_0,w'\}$. This is also true when $i_0=1$, since $w_0$ then has length $1$ and so $\sigma=w_0\in\{w_0,w'\}$.  When $\sigma=w'$,
\[
|p|=|s|=|w_1|+|\sigma|>|\sigma|=|w'|\geq|p|
\]
and so we have a contradiction. When $\sigma=w_0$, we see that $s=w_0w_1=w\in\{w,w'\}$. Hence the inductive step follows in this case.

Now assume that $|s|<|w_1|$. Since $s$ is non-empty, we see that $i_1=|w_1|\geq 2$ and hence $w_1\in X_{i_1}$. The statement $P(i_1+j)$ implies that $s\in\{w_1,w'\}$. Since $|s|<|w_1|$, we cannot have $s=w_1$, and so $s=w'\in\{w,w'\}$. Hence the inductive step follows in this case.

It remains to consider the case when $|s|=|w_1|$.

Suppose for a contradiction that $|p|=|w'_0|$. So $s=w_1$ and $p=w'_0$, which implies $w_1=w'_0$ and $i_1=j_0$. Now $s=w_1\in R_{i_1}$, and
$p=w'_0\in L_{j_0}=L_{i_1}$. But $R_{i_1}$ and $L_{i_1}$ are disjoint sets, so $s\not=p$ and we have a contradiction, as required.

Now suppose that $|p|>|w'_0|$. Then $w'_1$ has a non-empty prefix $\pi$ (of length $|p|-|w'_0|$) that is a suffix of $w$. When $j_1\geq 2$, since $w\in X_i$ and $w'_1\in X_{j_1}$, the statement $P(i+j_1)$ implies that $\pi\in\{w'_1,w\}$. This is also true when $j_1=1$, since $w'_1$ then has length $1$ and so $\pi=w'_1\in\{w,w'_1\}$.  When $\pi=w$,
\[
|s|=|p|=|w'_0|+|\pi|>|\pi|=|w|\geq |s|
\]
and so we have a contradiction. When $\pi=w'_1$, we see that $p=w'_0w'_1=w'\in\{w,w'\}$. Hence the inductive step follows in this case.

Finally, suppose that $|p|<|w'_0|$. Since $p$ is non-empty, we see that $j_0=|w'_0|\geq 2$ and hence $w'_0\in X_{j_0}$. The statement $P(i+j_0)$ implies that $p\in\{w,w'_0\}$. Since $|p|<|w'_0|$, we cannot have $p=w'_0$, and so $p=w\in\{w,w'\}$. Hence the inductive step follows in this case.

The theorem now follows by induction on $k$.
\end{proof} 

Set $L=\bigcup_{i=1}^{n-1} L_i$, and $R=\bigcup_{i=1}^{n-1} R_i$. Note that these expressions are in fact disjoint unions. Now, $L_i\cap R_i=\emptyset$ for all $i$ by definition, and clearly $L_i\cap R_j=\emptyset$ when $i\not=j$ as the elements of $L_i$ and $R_j$ have different lengths. So in fact the union
\[
L\cup R=\left(\bigcup_{i=1}^{n-1} L_i\right)\cup\left(\bigcup_{i=1}^{n-1} R_i\right)
\]
is disjoint. The definition of $L$ implies that $L$ is a set of prefixes of codewords in $X$. The definition of $R$ implies that $R$ contains no prefixes of codewords in $X$. Note that $X_n$ is exactly the set of words of length $n$ that are of the form $w_1w_2$ for $w_1\in L$ and $w_2\in R$.

For a codeword $w\in X$, an \emph{$\{L,R\}$-decomposition} is a list
\[
\underline{x}=(x_1,x_2,\ldots,x_k)
\]
of non-empty words $x_i$ such that $w$ can be expressed as the concatenation $w=x_1x_2\cdots x_k$ of the words $x_j$, and such that $x_j\in L\cup R$ for all $j$. We note that $x_1\in L$, since $x_1$ is a prefix of $w$. Moreover, $x_k\in R$, since $X$ is non-overlapping. Also we see that $k>1$, since all the words in $L$ and $R$ have length strictly less than $n$.

For a fixed codeword $w$, let $D_w$ be the set of $\{L,R\}$-decompositions of $w$. We define a rewriting system on $D_w$ as follows. Suppose $\underline{x}=(x_1,x_2,\ldots,x_k)\in D_w$ for some $k>2$ is such that $x_j\in L$ and $x_{j+1}\in R$ for some $j$ with $1\leq j<k$. Then $x_jx_{j+1}\in X_{|x_jx_{j+1}|}=L_{|x_jx_{j+1}|}\cup R_{|x_jx_{j+1}|}$, and so $x_jx_{j+1}\in L$ or $x_jx_{j+1}\in R$. Hence $\underline{y}=(x_1,\ldots,x_{j-1},x_jx_{j+1},x_{j+2},\ldots,x_k)$ is also an $\{L,R\}$-decomposition, and we write $\underline{x}\rightarrow\underline{y}$ for all transformations from $\underline{x}$ to $\underline{y}$ of this form. We write $\stackrel{*}{\rightarrow}$ for the transitive-reflexive closure of the rewriting system $\rightarrow$.

\begin{theorem}
\label{thm:rewriting}
Let $w=w_1w_2\cdots w_n\in X$ be fixed. The rewriting system $\stackrel{*}{\rightarrow}$ defined above is confluent, with a unique top element $\underline{t}=(w_1,w_2,\ldots,w_n)$ and a unique bottom element of the form $\underline{b}=(b_1,b_2)$, where $b_1\in L$ and $b_2\in R$.
\end{theorem}
\begin{proof}
We first show that $\rightarrow$ is locally confluent. Let $\underline{x},\underline{y},\underline{z}\in D_w$ be distinct and such that $\underline{x}\rightarrow \underline{y}$ and $\underline{x}\rightarrow\underline{z}$. Without loss of generality, there are integers $j,j'$ with $1\leq j<j'<k$ such that $\underline{y}$ is obtained from $\underline{x}$ by concatenating $x_j\in L$ and $x_{j+1}\in R$, and $\underline{z}$ is obtained from $\underline{x}$ by concatenating $x_{j'}\in L$ and $x_{j'+1}\in R$. Since $L\cap R=\emptyset$, we cannot have $j'=j+1$, and so the two concatenations involve disjoint pairs of words in $\underline{x}$. Define
\[
\underline{u}=(x_1,x_2,\ldots,x_{j-1},x_jx_{j+1},\ldots ,x_{j'}x_{j'+1},x_{j'+2},\ldots,x_k),
\]
so both concatenations are carried out to produce $\underline{u}$. Then $\underline{y}\rightarrow\underline{u}$ and $\underline{z}\rightarrow\underline{u}$. So $\rightarrow$ is locally confluent, as required.

We now show that $\stackrel{*}{\rightarrow}$ is confluent. First note that there are no infinite chains of non-identity rewriting operations, since each such operation reduces the number of words in the decomposition of $w$. The result now follows by Newman's theorem~\cite[Lemma~2.4]{Huet}, since $\rightarrow$ is locally confluent.

We show that the element $\underline{t}$ is the unique top element in $D_w$. Starting with an arbitrary element $\underline{x}\in D_w$, we may build a chain of the form
\[
\underline{t}\rightarrow\underline{t_1}\rightarrow\underline{t_2}\rightarrow\cdots\rightarrow\underline{x}
\]
by starting with $\underline{x}$, then repeating the following process: choose a term $x_j$ of length $m$ where $m>1$, writing $x_j=x'x''$ where $x'\in L$ and $x''\in R$, and then replacing $x_j$ by $x',x''$. (Note that $x_j\in X_m$, and so $x_j$ can indeed be written in the form $x'x''$ by the definition of $X_m$.) Reversing the sequence of decompositions obtained from this process produces a chain from $\underline{t}$ to $\underline{x}$. So, since every element lies below $\underline{t}$, we see that $\underline{t}$ is a top element in $D_w$. Indeed, $\underline{t}$ is the unique top element, since there is only one way of writing $w$ as the concatenation of alphabet symbols.

Confluence of $\stackrel{*}{\rightarrow}$, together with the fact that there is a unique top element, implies that the bottom element $\underline{b}$ of $\stackrel{*}{\rightarrow}$ is unique. Let $\underline{b}=(b_1,b_2,\ldots,b_k)$. We have already observed that $k>1$, $b_1\in L$ and $b_k\in R$. If $k\geq 3$, there exists an integer $j$ such that $1\leq j<k$, and both $b_j\in L$ and $b_{j+1}\in R$. But then we may rewrite $\underline{b}$ to produce a shorter decomposition, which contradicts the fact that $\underline{b}$ is a bottom element. Hence the final statement of the theorem follows. 
\end{proof}

\begin{corollary}
\label{cor:X_and_X_ell}
For any non-overlapping code $X$, we have $X\subseteq \overline{X}$. When $X$ is a maximal non-overlapping code, $X=\overline{X}$.
\end{corollary}
\begin{proof}
By Theorem~\ref{thm:rewriting}, every codeword $w\in X$ may be written as $w=b_1b_2$, where $b_1\in L$ and $b_2\in R$. Hence $w\in X_n$. So $X\subseteq \overline{X}$.

By Theorem~\ref{thm:balanced}, the pair $(X_n,X_n)$ is balanced, and so $\overline{X}=X_n$ is non-overlapping. When $X$ is maximal the fact that $X\subseteq X_n$ implies that $X=\overline{X}$.
\end{proof}

\begin{corollary}
\label{cor:unique_bottom}
Let $2\le i\le n$.
In the expression $X_i=\bigcup_{a\in\{1,2,\ldots,i-1\}}L_aR_{i-a}$, the union is disjoint.
\end{corollary}
\begin{proof}
Suppose $w\in L_aR_{i-a}\cap L_{a'}R_{i-a'}$ for some $a,a'\in\{1,2,\ldots ,i-1\}$ with $a<a'$. Write $w=u t v$, where $|u|=a$, $|t|=a'-a$ and $|v|=i-a'$. Then $ut$ belongs to $L_{a'}$ and $tv$ belongs to $R_{i-a}$. The nonempty word $t$ is a proper suffix of $ut$ and a proper prefix of $tv$. Both $ut\in X_{a'}$ and $tv\in X_{i-a}$ have length at least 2 and they share a common overlap $t$. This shows that the pair $(X_{a'},X_{i-a})$ is not balanced, contradicting Theorem~\ref{thm:balanced}. So the union is disjoint, as required.
\end{proof}

The following is a weaker form of~\cite[Proposition~10]{StanovnikMoskon24}:

\begin{theorem}
\label{thm:maximise}
Let $X$ be a maximal non-overlapping $q$-ary code of length $n$, where $n\geq 2$ and $q\geq 2$. Then there exist non-negative integers $\ell_1,\ell_2,\ldots ,\ell_{n-1}$ and $r_1,r_2,\ldots ,r_{n-1}$ with the following properties:
\begin{align}
\label{eqn:one}
\ell_1+r_1&=q\text{ where $\ell_1$ and $r_1$ are positive}\\
\label{eqn:other}
\ell_i+r_i&=\sum_{j=1}^{i-1} \ell_jr_{i-j}\text{ for $2\leq i\leq n-1$}\\
\label{eqn:cardinality}
|X|&=\sum_{i=1}^{n-1}\ell_ir_{n-i}.
\end{align}
\end{theorem}
\begin{proof}
Let $X$ be a maximal non-overlapping $q$-ary code of length $n$. Define sets $X_i$, $L_i$ and $R_i$ as above, and set $\ell_i=|L_i|$ and $r_i=|R_i|$ for $1\leq i\leq n-1$. Clearly the integers $\ell_i$ and $r_i$ are all non-negative.

Since $R_1=X_1\setminus L_1=F\setminus L_1$, we see that $\ell_1+r_1=q$. Now $X$ is a non-empty code, so the set $L_1$ of length $1$ prefixes of codewords must be non-empty. Hence $\ell_1$ is positive. If all $q$ length $1$ words occur as prefixes in $X$, then $X$ cannot be non-overlapping (as there is no valid choice for the end of a codeword that does not lead to an overlap of length $1$ between a pair of codewords). Hence $\ell_1\not=q$ and so $r_1$ is positive. Hence the condition~\eqref{eqn:one} holds.

For $2\leq i\leq n$ we have
\begin{equation}
\label{eqn:Xi_bound}
|X_{i}|=\left|\bigcup_{j=1}^{i-1}L_jR_{i-j}\right|= \sum_{j=1}^{i-1} \ell_jr_{i-j},
\end{equation}
as the union is disjoint by Corollary~\ref{cor:unique_bottom}.
In particular, when $2\leq i\leq n-1$, we have
\[
\ell_i+r_i=|L_i\cup R_i|=|X_i|= \sum_{j=1}^{i-1} \ell_jr_{i-j}.
\]
So the equalities~\eqref{eqn:other} hold.

Since $X$ is maximal, we have that $X=\overline{X}$ by Corollary~\ref{cor:X_and_X_ell}. Hence, by the definition of $\overline{X}$,
\[
|X|=|\overline{X}|=|X_n|=\sum_{j=1}^{n-1} \ell_jr_{n-j},
\]
the last equality following by~\eqref{eqn:Xi_bound}. Hence~\eqref{eqn:cardinality} holds, and the theorem holds.
\end{proof}

The theorem allows us to provide an upper bound on the cardinality of a non-overlapping code of length $n$, by an integer maximisation problem.

\section{An improved upper bound}
\label{sec:improved_upper}

This section provides a proof of Theorem~\ref{thm:main}, using Theorem~\ref{thm:maximise}. The section is structured as follows. Subsection~\ref{sub:intro} provides an introduction. We rephrase the problem in a form where we can apply generating function theory, we rederive Levenshtein's upper bound, and we point out where there is room for the bound to be improved. Subsection~\ref{sub:setting_up} rephrases the problem, defining notation to avoid writing many powers of $q$ in the rest of the section, and preparing for the bounds we describe later. Subsection~\ref{sub:family} provides a family of bounds on a non-overlapping code, a bound for each integer $d$ with $1\leq d<n$. The bound consists of a sum of two terms, the one of which will give us the correct constant in our bound. In Subsection~\ref{sub:end_main_proof} we show that, when $d$ is chosen appropriately, the second term in this bound is negligible. This will give us the upper bound we require. 

\subsection{Introduction}
\label{sub:intro}

In this subsection we are only interested in binary codes and so we set $q=2$ throughout.

We let $X$ be a maximal binary non-overlapping code with $|X|=m$, where $m$ is as large as possible. Using Theorem~\ref{thm:maximise}, we let $\ell_i$ and $r_i$ be non-negative integers satisfying~\eqref{eqn:one} and~\eqref{eqn:other}, such that $m= \left(\sum_{i=1}^{n-1}\ell_i r_{n-i} \right)$. How do we find an upper bound on $m$?

Define real polynomials $L(z)$ and $R(z)$ in the variable $z$ by
\[
L(z)=\sum_{i=1}^{n-1}\ell_iz^i\text{ and }R(z)=\sum_{i=1}^{n-1}r_iz^i.
\]
The definition of $m$, together with the relations~\eqref{eqn:one} and~\eqref{eqn:other}, shows that
\begin{equation}
\label{eqn:polynomials}
(1-L(z))(1-R(z))=1-2z+mz^n+T(z),
\end{equation}
where $T(z)$ is a polynomial of degree at most $2n-2$ which is divisible by $z^{n+1}$. Indeed, writing
\begin{equation}
\label{eqn:S}
S=\{(i,j):1\leq i,j\leq n-1\text{ and }i+j>n\}
\end{equation}
we see that
\[
T(z)=\sum_{(i,j)\in S}\ell_ir_jz^{i+j}=\sum_{k=n+1}^{2n-2}c_kz^k
\]
for some non-negative coefficients $c_k$.

The polynomial $1-L(z)$ evaluates to $1$ at $z=0$, and has a negative leading coefficient. So $1-L(z)\rightarrow -\infty$ as $z\rightarrow\infty$. Hence $1-L(z)$ has a root on the positive real axis. Similarly, $1-R(z)$ has a root on the positive real axis. Let $\rho$ be the smallest positive real number such that $\rho$ is a root of one of $1-L(z)$ and $1-R(z)$. Evaluating~\eqref{eqn:polynomials} at $\rho$, we see that
\begin{equation}
\label{eqn:evaluation}
0=1-2\rho+m\rho^n+T(\rho).
\end{equation}

Since $\rho$ is positive and all the coefficients of the non-zero polynomial $mz^n+T(z)$ are non-negative, we see that $m\rho^n+T(\rho)>0$ and so~\eqref{eqn:evaluation} implies that $\rho>1/2$.

At this point, we can rederive Levenshtein's upper bound on the size of a binary non-overlapping code, as follows. Since the coefficients of $T(z)$ are non-negative, and $\rho$  is positive, we have $T(\rho)\geq 0$ and so~\eqref{eqn:evaluation} implies that $2\rho-1\geq m\rho^n$. Hence
\[
m\leq \frac{2\rho-1}{\rho^n}.
\]
Using standard calculus, we can maximise the function $f(t)=(2t-1)/t^n$ when $t>1/2$. We find a single maximum when $t=t^*:=n/(2n-2)$, at the value
\[
f(t^*)=\frac{2^n}{n}\left(\frac{n-1}{n}\right)^{n-1}.
\]
Hence
\[
m\leq f(t^*)=\frac{2^n}{n}\left(\frac{n-1}{n}\right)^{n-1}=\left(\frac{1}{e}+O(n^{-1})\right)\frac{2^n}{n}.
\]

How can we improve this bound? We used the fact that $T(\rho)\geq 0$ above. A sufficiently large lower bound for $T(\rho)$ would improve this estimate. But instead we truncate both generating polynomials $L(x)$ and $R(x)$ and compare the degree-$n$ coefficient of the full product with a weighted sum of coefficients of the truncated product. This allows us to improve the constant in our upper bound from $1/e$ to $1/(2e)$ in the binary case (and we show something similar works for arbitrary finite alphabets).

\subsection{Setting up the problem}
\label{sub:setting_up}

We return to the general $q$-ary case, so we have $q\geq 2$.
Let $X$ be a maximal $q$-ary non-overlapping code of length $n$. Define the integers $\ell_i$ and $r_i$ as in Theorem~\ref{thm:maximise}. Write $m=\sum_{i=1}^{n-1}\ell_i r_{n-i}$, so $|X|=m$. 
To avoid writing many powers of $q$, we reparametrise, writing $a_i=q^{-i}\ell_i$, $b_i=q^{-i}r_i$, and 
\[
 p=\sum_{i=1}^{n-1}a_i b_{n-i}
   =q^{-n}\sum_{i=1}^{n-1}\ell_i r_{n-i}.
\]
We note that $p=mq^{-n}$, and so Theorem~\ref{thm:main} follows if we can show that $np\leq (q-1)/(qe)+O(n^{-1/4})$.

The equalities~\eqref{eqn:one} and~\eqref{eqn:other} imply that $a_1+b_1=1$, that $a_1,b_1\ge1/q$, and that
\begin{equation}
\label{eqn:a_plus_b}
a_i+b_i=\sum_{j=1}^{i-1}a_j b_{i-j}\text{ for } 2\le i<n.
\end{equation}
For \(1\le d<n\), we define
\[
 A_d(z)=\sum_{i=1}^{d}a_i z^i \text{ and }B_d(z)=\sum_{i=1}^{d}b_i z^i,
 \]
 and we define the polynomial $P_d(z)$ by the equality
 \begin{equation}
 \label{eqn:P_def}
 (1-A_d(z))(1-B_d(z))=1-z+P_d(z).
\end{equation}
By~\eqref{eqn:one} and~\eqref{eqn:other}, the lowest $d+1$ coefficients of $P_d(z)$ are all zero, and indeed
 \(P_d\) consists of those terms in the product \(A_dB_d\) of
degree at least \(d+1\). In particular all the coefficients of $P_d$ are
non-negative.

Write \(A=A_{n-1}\) and \(B=B_{n-1}\). Thus, writing $\coeff{n}Q$ for the coefficient of $z^n$ of some polynomial or power series $Q(z)$, we have
\(\coeff{n}P_{n-1}=p\).

For each integer \(s\ge2\), define \(\lambda_s=s^{-1}(1-1/s)^{s-1}\). It is clear that $\lambda_s<1/s$. Moreover, when $s\rightarrow\infty$ we see that 
\begin{equation}
\label{eqn:lambda_getting_big}
s\lambda_s=(1/e)+O(s^{-1}).
\end{equation}

\begin{lemma}\label{lem:prefix}
For every \(1\le d<n\), we have
\begin{equation}
\label{eqn:AB_one}
 A_d(1)\leq 1\text{ and }B_d(1)\le1.
 \end{equation}
Moreover,
\begin{equation}\label{eq:prefix}
 (1-A_d(1))(1-B_d(1))=P_d(1)\le\lambda_{d+1}.
\end{equation}
\end{lemma}

\begin{proof}
Let \(\tau\) be the smaller of the positive solutions of \(A_d(\tau)=1\)
and \(B_d(\tau)=1\). (These solutions exist and are unique since both
polynomials are strictly increasing from zero to infinity on the
positive real axis.) Evaluating~\eqref{eqn:P_def} at $\tau$ shows that \(P_d(\tau)=\tau-1\). The coefficients of $P_d$ are non-negative, therefore $P_d$ is non-negative on the positive real axis, and so  \(\tau\ge1\). Hence \(A_d(1),B_d(1)\le1\), and we have shown \eqref{eqn:AB_one} holds.

The first equality in~\eqref{eq:prefix} holds by evaluating~\eqref{eqn:P_def} at $1$.

Now \(P_d\) is divisible by \(z^{d+1}\). Using the fact that $P_d(z)/z^{d+1}$ is non-decreasing on the positive real numbers,
\[
 P_d(1)=1^{-(d+1)}P_d(1) \le \tau^{-(d+1)}P_d(\tau)=\frac{\tau-1}{\tau^{d+1}}\leq\frac{t-1}{t^{d+1}} ,
\]
where $t$ is the real number which maximises the function $(x-1)x^{-(d+1)}$ over $x\geq 1$. The derivative of \((x-1)x^{-s}\) is
\(x^{-s-1}(s-(s-1)x)\), so its maximum over \(x\ge1\) occurs at
\(t=s/(s-1)\) and equals \(\lambda_s\). Taking \(s=d+1\) gives the inequality in~\eqref{eq:prefix}, and the lemma follows.
\end{proof}

We shall need two consequences of Lemma~\ref{lem:prefix}. First, for \(2\le i<n\), we have
\begin{equation}\label{eq:point}
 a_i+b_i=\coeff{i}P_{i-1}\le P_{i-1}(1)\le\lambda_i\le\frac1i .
\end{equation}
To see this, note that the equality follows by~\eqref{eqn:a_plus_b} and the definition of $P_{i-1}$; the first inequality follows from the fact the coefficients of $P_{i-1}$ are non-negative; the next inequality follows from Lemma~\ref{lem:prefix}; the final inequality follows from the definition of $\lambda_i$.
Secondly, for \(1\le k<n\), we have 
\begin{equation}\label{eq:tails}
 \left(\sum_{i=k+1}^{n-1}a_i\right)
 \left(\sum_{i=k+1}^{n-1}b_i\right)
 \le (1-A_k(1))(1-B_k(1))
 \le\lambda_{k+1}\le\frac1{k+1}.
\end{equation}
To see this, note that the first inequality uses the fact that the sums are equal to $A(1)-A_k(1)$ and $B(1)-B_k(1)$ respectively, together with the fact that \(A(1),B(1)\le1\) by Lemma~\ref{lem:prefix} with \(d=n-1\). As before, the other inequalities follow from Lemma~\ref{lem:prefix} and the definition of $\lambda$.

\subsection{A family of bounds on $p$}
\label{sub:family}

The following lemma contains the key argument which gives us the improved factor $(q-1)/q$ in our upper bound. Recall that we need to provide an upper bound on $p=\coeff{n}P_{n-1}$. For $d<n-1$, the definition of $P_d$ shows that we have $\coeff{n}P_d\leq \coeff{n}P_{n-1}$, and so $\coeff{n}P_d$ does not provide an upper bound on $p$. The remarkable fact is that a weighted sum (encoded by the power series $W$ in the proof below) of the coefficients of $P_d$ \emph{does} provide such an upper bound.

\begin{lemma}\label{lem:half}
For every \(1\le d<n\),
\begin{equation}\label{eq:half}
 p\le (1-1/q)P_d(1)+q^{-1}\coeff{n}P_d.
\end{equation}
\end{lemma}

\begin{proof}
We work with formal power series, and introduce
\[
 W(z)=\frac{1-z/q}{1-z}
     =1+(1-1/q)\sum_{h\ge1}z^h .
\]
We claim that both of the products \(W(1-A)\) and \(W(1-B)\) have non-negative coefficients
up to (and including) degree \(n-1\). To see this, note that \((1-A)(1-B)\equiv 1-z\pmod{z^n}\),
so
\begin{equation}
\label{eqn:W_product}
 W(1-A)\equiv\frac{1-z/q}{1-B}
   =1+\frac{B-z/q}{1-B}\pmod{z^n}.
\end{equation}
Here \(B-z/q\) has non-negative coefficients since \(b_1\ge1/q\),
and \((1-B)^{-1}=\sum_{j\ge0}B^j\) has non-negative coefficients too. Hence \(W(1-A)\) has non-negative coefficients up to degree $n-1$ as claimed. Interchanging \(A\) and \(B\) in the argument above shows that \(W(1-B)\) has non-negative coefficients up to degree $n-1$, and the claim follows.

Put \(U=A-A_d\) and \(V=B-B_d\). A short calculation, using the definitions of $U$, $V$ and the definition~\eqref{eqn:P_def}, shows that $P_d-P_{n-1}=U(1-B)+V(1-A)+UV$, and so
\begin{equation}\label{eq:identity}
 W(P_d-P_{n-1})
   = U\,W(1-B)+V\,W(1-A)+WUV.
\end{equation}

We claim that the coefficient of \(z^n\) in~\eqref{eq:identity} is non-negative. To see this, we examine the right hand side and show that the coefficient of $z^n$ in each summand is non-negative. The last summand has all coefficients non-negative, since the same is true for each of the factors $U$, $V$ and $W$. For the other two terms, note that
\(U,V\) have non-negative coefficients and zero constant term, so
only coefficients of \(W(1-B)\) and \(W(1-A)\) of degree below \(n\)
are involved in the coefficient of $z^n$, and so our claim follows by the conclusion of the previous paragraph. Rephrasing, we have shown that $\coeff{n}(WP_{n-1})\leq \coeff{n}(WP_d)$.

Since \(P_{n-1}\) is divisible by $z^n$, and since $W$ is a power series with constant term $1$ and all other coefficients equal to $1-1/q$, we calculate
\begin{align*}
 p&=\coeff{n}(WP_{n-1})\le \coeff{n}(WP_d)
   =\coeff{n}P_d+(1-1/q)\sum_{h=0}^{n-1}\coeff{h}P_d\\
   &\le(1-1/q)P_d(1)+q^{-1}\coeff{n}P_d.
 \qedhere
\end{align*}
\end{proof}

\subsection{The proof of Theorem~\ref{thm:main}}
\label{sub:end_main_proof}

Lemma~\ref{lem:half} shows that we can use $P_d$ to bound $p$. We are free to choose~$d$ to give as sharp a bound as possible. We will choose $d$ so that the second term in the lemma is small, but the first term still gives us a tight bound.

Suppose \(1\le k\le(n-2)/2\) and put \(d=n-k-1\). Every term $a_ib_{n-i}$ in the sum defining \(\coeff{n}P_d\) has both indices in
\(I=\{k+1,\ldots,n-k-1\}\). For \(i\in I\), the equation~\eqref{eq:point} gives
\[
 a_i b_{n-i}\le(a_i+b_i)(a_{n-i}+b_{n-i})\leq \frac1{i(n-i)}
           \le\frac1{(k+1)(n-k-1)}.
\]
We bound one factor of
\(a_i b_{n-i}=\sqrt{a_i b_{n-i}}\,\sqrt{a_i b_{n-i}}\)
using this inequality. Summing the remaining square root factor over $i\in I$ and applying the
Cauchy--Schwarz inequality gives
\begin{align*}
 \coeff{n}P_d
  &=\sum_{i\in I}a_i b_{n-i}\\
  &\le\frac{1}{\sqrt{(k+1)(n-k-1)}}
     \left(\sum_{i\in I} a_i\right)^{1/2}
     \left(\sum_{i\in I} b_{n-i}\right)^{1/2}\\
  &\le\frac{1}{(k+1)\sqrt{n-k-1}},
\end{align*}
where the final step follows from \eqref{eq:tails}, since \(n-i\in I\)
whenever \(i\in I\). By \eqref{eq:prefix} and \eqref{eq:half} we
have proved the finite inequality
\begin{equation}\label{eq:finite}
 p\le(1-1/q)\lambda_{n-k}
              +\frac1{q(k+1)\sqrt{n-k-1}}.
\end{equation}

We take \(k=\lfloor n^{3/4}\rfloor\), which lies in the required range for
all sufficiently large \(n\). Since $n-k\rightarrow \infty$, we have $(n-k)\lambda_{n-k}= (1/e)+O((n-k)^{-1})$, by~\eqref{eqn:lambda_getting_big}, and so
\[
 n\lambda_{n-k}=\frac1e+O(n^{-1/4})
 \]
 Moreover,
 \[
 \frac{n}{(k+1)\sqrt{n-k-1}}=O(n^{-1/4}).
\]
Multiplying \eqref{eq:finite} by \(n\) and applying these last two asymptotic formulas proves Theorem~\ref{thm:main}.

\section*{Acknowledgements} This paper was partly written on a visit to Tuvi Etzion at Technion, Israel, in October and November 2022. My thanks to Tuvi and to Technion for their great hospitality during my stay. An earlier, weaker, version of the proof in Section~\ref{sec:improved_upper} was generated by OpenAI GPT-6 Astra; the full result was then generated by using GPT-6.2 Astra. The author has independently verified (and takes responsibility for) this proof. The proof in Section~\ref{sec:improved_upper} was drafted by GPT-6.2 Astra, and edited for clarity by the author. All other writing is due to the author.


\begin{thebibliography}{99}
\bibitem{Bajic} Dragana Baji\'c, ``On construction of cross-bifix-free
  kernel sets'', presented at 2nd COST 2100 MCM, Lisbon, Portugal, February 2007.
\bibitem{BajicStojanovic} Dragana Baji\'c and Jakov Stojanovi\'c,
  ``Distributed sequences and search process'', \emph{Proc. IEEE
  Internat. Conf. on Communications}, pp. 514--518, 2004.
\bibitem{BajicStojanovicLindner} Dragana Baji\'c, Jakov Stojanovi\'c
  and Juergen Lindner, ``Multiple window-sliding search'',
  \emph{Proc. Int. Symp. Inf. Theory}, p. 249, 2003.
\bibitem{BilottaGrazzini} Stefano Bilotta, Elisabetta Grazzini, Elisa
  Pergola and Renzo Pinzani, ``Avoiding cross-bifix-free binary words'',
  \emph{Acta Informatica}, vol. 50, pp.~157--173, 2013.
\bibitem{BilottaPergola} Stefano Bilotta, Elisa Pergola and Renzo
  Pinzani, ``A new approach to cross-bifix-free sets'', \emph{IEEE
    Trans. Inf. Theory}, vol. 58, pp. 4058--4063, 2012.
 \bibitem{Blackburn15} Simon R.\  Blackburn, ``Non-overlapping codes'', \emph{IEEE Trans. Inform. Theory}, vol. 61, pp. 4890--4894, 2015. 
\bibitem{CheeKiah13} Yeow Meng Chee, Han Mao Kiah, Punarbasu
  Purkayastha and Chengmin Wang, ``Cross-bifix-free codes within a
  constant factor of optimality'', \emph{IEEE Trans. Inf. Theory}, vol. 59, pp. 4668--4674, 2013.
\bibitem{FimmelMichel19} Elena Fimmel, Christian Michel, François Pirot, Jean-Sébastien Sereni, Lutz
Str\"{u}ngmann, ``Comma-free Codes Over Finite Alphabets'', \emph{HAL Archives Ouvertes}, \url{https://hal.archives-ouvertes.fr/hal-02376793}, submitted 22nd November 2019.
\bibitem{FimmelMichel23} Elena Fimmel, Christian J. Michel, François Pirot, Jean-Sébastien Sereni, Lutz
Str\"{u}ngmann, ``Diletter and triletter comma-free codes over finite alphabets'', \emph{Australas. J. Combin.}, vol. 86, 233--270, 2023.
\bibitem{Gilbert60} E.N. Gilbert, ``Synchronization of binary messages'',
  \emph{IRE Trans. Inf. Theory}, vol. 6, pp. 470--477, 1960.
\bibitem{Huet} G\'{e}rard Huet, `Confluent reductions: abstract properties and applications to term rewriting systems', \emph{J. ACM} \textbf{27} (1980), 797--821.
\bibitem{Levenshtein64} V.I. Levenshtein, ``Decoding automata which are invariant with respect to their initial state'', \emph{Probl. Cybern.}, vol. 12, pp. 125--136, 1964 (in Russian).
\bibitem{Levenshtein70} V.I. Levenshtein, ``Maximum number of words in codes without overlaps'', \emph{Prob. Inform. Transmission}, vol. 6, pp. 355--357, 1970.
\bibitem{Levenshtein04} V.I. Levenshtein, ``Combinatorial problems
  motivated by comma-free codes'', \emph{J. Combinatorial Designs}, vol. 12, pp. 184--196, 2004.
\bibitem{WijngaardenWillink} Adriaan J. de Lind van Wijngaarden and
  Tricia J. Willink, ``Frame synchronization using distributed
  sequences'', \emph{IEEE Trans. Comm.}, vol. 48,
  pp. 2127--2138, 2000.
 \bibitem{StanovnikMoskon24} Lidija Stanovnik, Miha Mo\v skon and Miha Mraz, ``In search of maximum non-overlapping codes'', \emph{Designs, Codes and Cryptography}, vol. 92, 1299--1326, 2024. 
 \bibitem{Zakharov26} Dmitrii Zakharov, ``An isoperimetric inequality for word overlap'', \emph{arXiv}, \url{https://arxiv.org/abs/2602.20143}, 3 March 2026.
\end{thebibliography}
\end{document}